\documentclass[letterpaper,amsmath,amssymb,preprint,superscriptaddress]{revtex4}
\usepackage{amstext}
\usepackage{graphicx} 
\usepackage{dcolumn}   
\usepackage{bm}      
\usepackage{amsmath}  
\usepackage{indentfirst} 
\usepackage{psfrag}      
\usepackage{subfigure}   
\usepackage{amssymb}     
\usepackage[colorlinks,linkcolor=blue,citecolor=blue,urlcolor=blue,hyperindex]{hyperref}

\begin{document}
\title{Active plasmon interference control on ultrafast time scales by free electrons}

\author{Rong Huang}
\affiliation{The Key Laboratory of Weak-Light Nonlinear Photonics, Ministry of Education, School of Physics and TEDA Applied Physics Institute, Nankai University, Tianjin 300457, China}

\author{Weiwei Luo}
\affiliation{The Key Laboratory of Weak-Light Nonlinear Photonics, Ministry of Education, School of Physics and TEDA Applied Physics Institute, Nankai University, Tianjin 300457, China}

\author{Wei Wu}
\affiliation{The Key Laboratory of Weak-Light Nonlinear Photonics, Ministry of Education, School of Physics and TEDA Applied Physics Institute, Nankai University, Tianjin 300457, China}

\author{Mengxin  Ren}
\affiliation{The Key Laboratory of Weak-Light Nonlinear Photonics, Ministry of Education, School of Physics and TEDA Applied Physics Institute, Nankai University, Tianjin 300457, China}

\author{Xinzheng Zhang}
\affiliation{The Key Laboratory of Weak-Light Nonlinear Photonics, Ministry of Education, School of Physics and TEDA Applied Physics Institute, Nankai University, Tianjin 300457, China}

\author{Wei Cai}
\email{weicai@nankai.edu.cn}
\affiliation{The Key Laboratory of Weak-Light Nonlinear Photonics, Ministry of Education, School of Physics and TEDA Applied Physics Institute, Nankai University, Tianjin 300457, China}
\affiliation{Collaborative Innovation Center of Extreme Optics, Shanxi University, Taiyuan, Shanxi
030006, People's Republic of China}

\author{Jingjun Xu}
\email{jjxu@nankai.edu.cn}
\affiliation{The Key Laboratory of Weak-Light Nonlinear Photonics, Ministry of Education, School of Physics and TEDA Applied Physics Institute, Nankai University, Tianjin 300457, China}
\date{\today}

\begin{abstract}
Interference between light waves is one of the widely known phenomena in physics, which is widely used in modern optics, ranging from precise detection at the nanoscale to gravitational-wave observation. Akin to light, both classical and quantum interferences between surface plasmon polaritons (SPPs) have been demonstrated. However, to actively control the SPP interference within subcycle in time (usually less than several femtoseconds in the visible range) is still missing, which hinders the ultimate manipulation of SPP interference on ultrafast time scale.  In this paper, the interference between SPPs launched by a hole dimer, which was excited by a grazing incident free electron beam without direct contact, was manipulated through both propagation and initial phase difference control. Particularly, using cathodoluminescence spectroscopy, the appearance of higher-order interference orders was obtained through propagation phase control by increasing separation distances of the dimer.  Meanwhile, the peak-valley-peak evolution at a certain wavelength through changing the accelerating voltages was observed, which originates from the initial phase difference control of hole launched SPPs. In particular, the time resolution of this kind of control is shown to be in the ultrafast attosecond (as) region. Our work suggests that fast electron beams can be an efficient tool to control polarition interference in subcycle temporal scale, which can be potentially used in ultrafast optical processing or sensing.
\end{abstract}

\maketitle

\section{INTRODUCTION}
Interference is one of the most important physical phenomena for light, and is widely used in modern optical science and technology. Meanwhile, surface plasmon polaritons (SPPs) are the collective oscillations of conduction electrons coupled with light, propagating along a metal-dielectric interface \cite{barnes2003surface}. Because of their half light property, SPPs possess lots of similar properties like light and can be recognized as a kind of photonic quasiparticles. The interference of SPPs plays a vital role in nano-optics. In specific, the interference of SPPs provides an efficient way to control light propagation and light-matter interaction at the nanoscale. It has been extensively explored in many applications, such as optical logic gates\cite{wei2011cascaded,fu2012all}, biosensors\cite{gao2013plasmonic}, optical switches\cite{bahrami2011all,sahu2015theoretical,choi2017control}, and nanolithography\cite{luo2004surface,liu2005surface}. Besides, the plasmonic version of quantum interference has been demonstrated as well, in which single \cite{DDE2016} and two-plasmon \cite {HK2013, FLK2014, VDD2017} quantum interference were reported and can be potentially used in quantum plasmonic circuits. 

Up to now, interference between SPPs has been investigated very extensively under different schemes. For example, Young's double-slit is widely used to acquire the interference patterns and modulate the visibility of the interference fringes by varying the slit-slit separation\cite{gan2007surface,li2017strong} or changing one of the slits into groove\cite{morrill2016measuring}.  Moreover, propagating SPPs in metal films can be focused into a bright spot by periodic surface defects and plasmonic lens due to the constructive interference\cite{liu2005focusing, lopez2007efficient, lerman2009demonstration, chen2010experimental}. The near-field interference patterns can be controlled by delicate illumination design, as well as by structural tuning \cite{dvorak2013control}. Besides, the interference between SPPs with light can also be realized, such as the interaction between SPPs and electron excited transition radiation \cite{kuttge2009local}. Yet, in most of the previous studies, the interference of SPPs is controlled by tuning the propagation phase of plasmons. As we know, except for the propagation phase of waves, the initial phase difference is also critical for the interference effect. It is still a lack of consideration of the initial phase carried by SPPs. In parallel, free electron beams have been demonstrated as an effect plasmon source at the nanoscale \cite{cai2009efficient, abajo2010}. Particularly, electron beam based cathodoluminiscence (CL) spectroscopy can excite SPPs with nanoscale resolution through direct bombardment on the sample, and the optical radiation emitted from the sample is detected\cite{vesseur2008surface,knight2012aluminum,du2018evolution,li2018cathodoluminescence}. Furthermore,  an electron beam that passes parallel to the surface of structures without direct contact can also excite SPPs \cite{yamamoto2001photon}. As a result, if one can set several plasmon sources along the path of a nearby electron beam, the initial phase difference of excited SPPs by these sources can be designed. More importantly, the initial phase difference and resulted time delay depends not only on the spatial distribution of the plasmon sources, but also on the speed of the electrons, providing an active way to control the interference.

In this paper, nonohole dimers were chosen as sources for SPPs. By using CL spectroscopy with grazing incident electrons, the interference of SPPs launched by nanoholes was investigated not only by propagating phase control through hole separation distance tuning, but also by manipulating the initial phase difference of SPPs excited at the two individual nanoholes by electrons.  Such manipulation is shown to be on ultrafast attosecond time scale, which is determined by the separation distance between two holes and the velocity of the electron beam, providing an efficient method to control SPP interference within one period of SPPs. Meanwhile, full-wave electromagnetic simulations based on the Finite-difference time-domain (FDTD) method reproduced well with the experimental measured data.

\section{Experiment and Results}
\begin{figure}[htbp]
\centering
\includegraphics[width=160mm,angle=0]{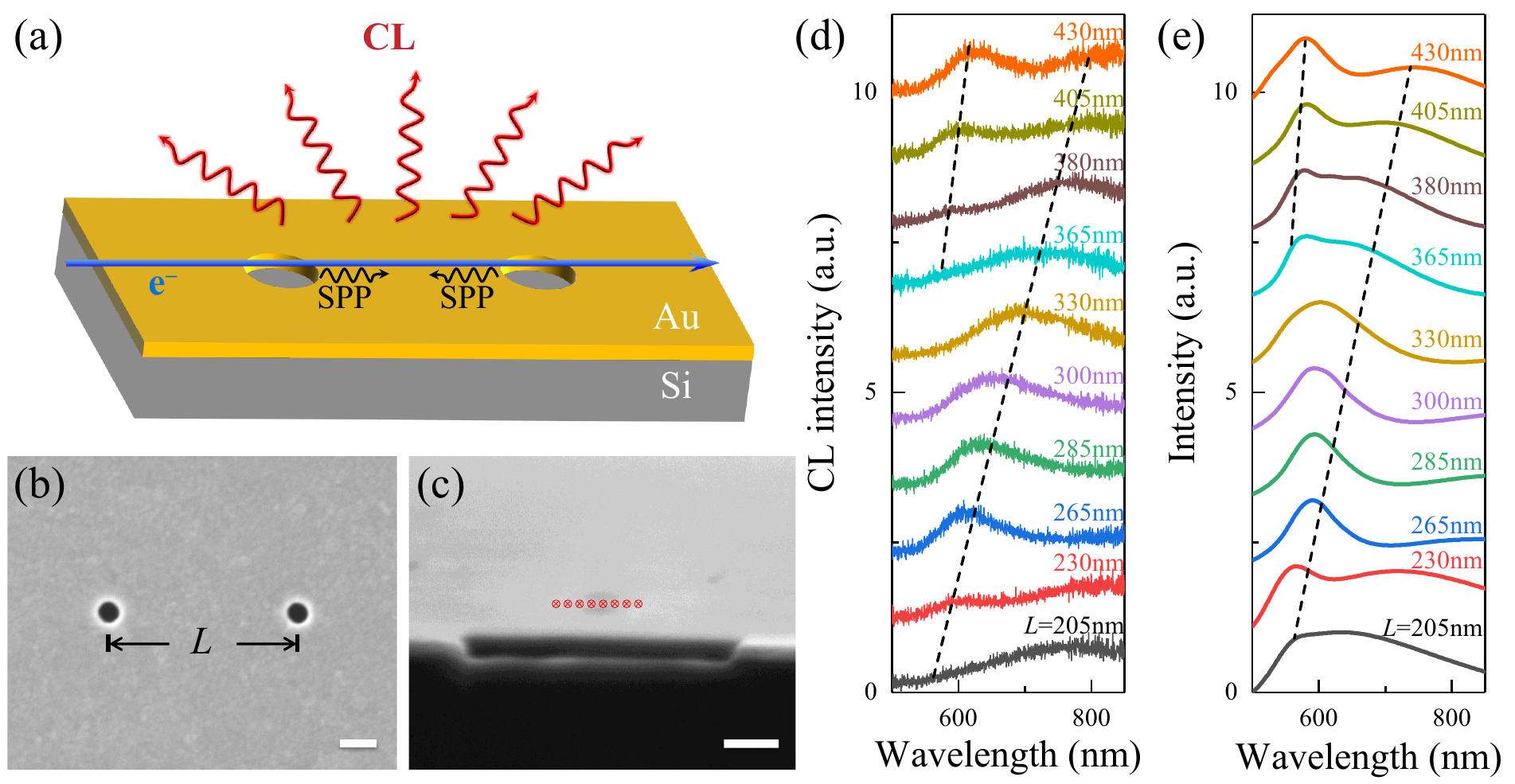}
\caption{(a) Schematic illustration of a nanohole dimer milled in a gold film on a silicon substrate. An electron beam is used to excite surface plasmon polaritons (SPPs) by the nanoholes in sequence under grazing incidence. The emission light induced by the interaction between SPPs is collected by the CL system. (b) Top-view SEM image of the nanohole dimer with a distance of $L=1000$~nm,  and the radius of each hole is $r=55$~nm. (c) SEM image of the 87$^\circ$-tilted sample. The electron beam was scanned along the red dashed line in the experiments. Scale bars, 200~nm. Experimental (d) and simulated (e) CL spectra of the nanohole dimers with nanohole separation distance $L=205-430$~nm.} \label{fig1}
\end{figure}

The schematic illustration of the experiments is shown in Fig.~\ref{fig1}(a). Particularly, an electron beam was used to interact with a nanohole dimer under grazing incidence (\emph{i.e.}, aloof configurations \cite{xxx}). The individual nanoholes in the dimer play the role of SPP source. Due to the limited speed of the electron beam, the holes interact with the beam in sequence. So the initial phases of the exctied SPPs are supposed to be different. As a result, the interference between the excited SPPs by these two holes can be manipulated.  In the experiments, the nanohole dimers were obtained in a 70~nm thick gold film on a silicon substrate using focused ion beam (FIB) milling, with a radius of each hole $r$=55~nm. The separation distance between the holes is defined by $L$, which was varied in the experiments. Figure~\ref{fig1}(b) and (c) show the top and 87$^\circ$-tilted view scanning electron microscopy (SEM) images of one typical dimer with $L$=1000~nm. The nanohole dimers were excited by the electron beam scanned along the red line with the impact parameter around 10~nm (Fig.~\ref{fig1}c). It is worth mentioning that the CL emission intensity is at its maximum when the electron beam passes along the symmetric axis of the dimer, but the resonance wavelength keeps almost unchanged with the different parallel trajectories (see Fig.~S1 in supplementary). As a result, the spectrum with the largest intensity in the line-scanning results for each nanohole dimer is chosen in the whole paper. Figure \ref{fig1}(d) shows CL spectra for nanohole dimers with the separation distance $L$ ranging from 205 to 430~nm. As the separation distance increases, the peak wavelength shifts to red, and a new peak starts to appear when the separation is 365~nm. The corresponding full-wave electromagnetic simulations using FDTD are shown in Fig.~\ref{fig1}(e), which agree well with the measured data. 

\begin{figure}[htbp]
\includegraphics[width=160mm,angle=0]{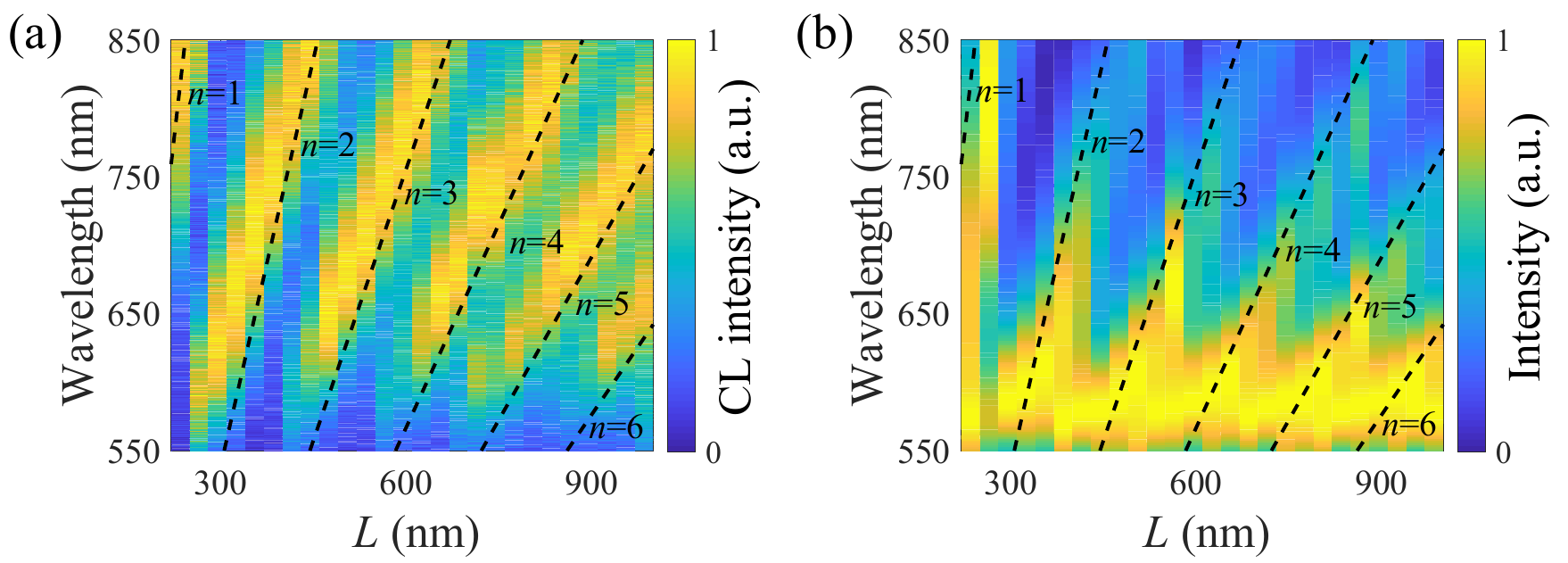}
\caption{ SPP interference control by propagation phase difference design. (a) Wavelength and nanohole separation distance resolved CL spectra of the nanohole dimers. The dashed lines indicate analytic results of constructive interference between SPPs excited by two holes, where $n$ represents the interference orders. (b) The corresponding full-wave electromagnetic simulations with (a).}\label{fig2}
\end{figure}

As we know, the plasmon response of nanohole dimers depends a lot on the separation distance between the holes. Two different interaction regions can be categorized. In the large separation limit, where the separation is far larger than the SPP wavelength, the SPPs launched by each hole will counterpropagate and then interact. On the contrary, the localized plasmon resonance of the dimer through near-field coupling will dominate as long as the separation distance is small enough. We first consider the former case.  An analytic model based on SPP interference is proposed to understand the physical mechanism behind it. Because the electron beam interacts with the holes of the dimer in sequence, the excited SPPs by the holes will have retardation in time ($L/v$), which depends on the separation distance and the velocity of the electron beams $v$. As a result,  there is an initial phase difference ($\omega L/v$) between the SPPs launched by the holes. In further, the total phase difference $\Delta\phi$ between the two conterpropagating SPP waves, including both propagation and initial phase differences, is determined by
\begin{equation}
\Delta\phi=k_{\text{spp}}(L-2r)+\frac{\omega L}{v}=\frac{2\pi}{\lambda_\text{spp}}(L-2r)+\frac{2\pi c}{\lambda_0}\frac{L}{v}
\label{eqn1}
\end{equation}
where $k_\text{spp}$ is the wave vector of SPPs, $c$ is the speed of light in vacuum, $\lambda_0$ is the wavelength of the emitted light in free space, and $\lambda_{\text{spp}}=\lambda_0\sqrt{1+1/\epsilon}$ is the SPP wavelength. $\epsilon$ is the dielectric constant of gold taken from Johnson and Christy \cite{johnson1972optical}. From this equation, constructive and destructive interference between two SPPs can be realized as long as $\Delta\phi=2n\pi$ or $(2n+1)\pi$ is fulfilled,  with the interference order $n$=$0, \pm1,\pm2,\pm3, \cdots$. On the experimental side, peaks and valleys in CL spectra should appear and correspond to constructive and destructive interference of SPPs respectively. Besides, one can know that the total phase difference can be tuned not only by the widely used propagation phase design with separation distance, but also by the velocity of the electron beam according to Eq.~\ref{eqn1}. Figure~\ref{fig2}(a) and (b) show the separation distance $L$ and wavelength resolved CL spectra from experiments and simulations, respectively.  For the same emission wavelength, the CL intensity oscillates as the variation of separation distance $L$, which hinders the possible interference of SPPs launched by the two holes. Therefore, analytical fits of constructive interference by using Eq.~\ref{eqn1} are plotted in dashed lines. Both the measured and simulated data are well reproduced with the analytical results with $n$=1-6. 

\begin{figure}[htbp]
\includegraphics[width=160mm,angle=0]{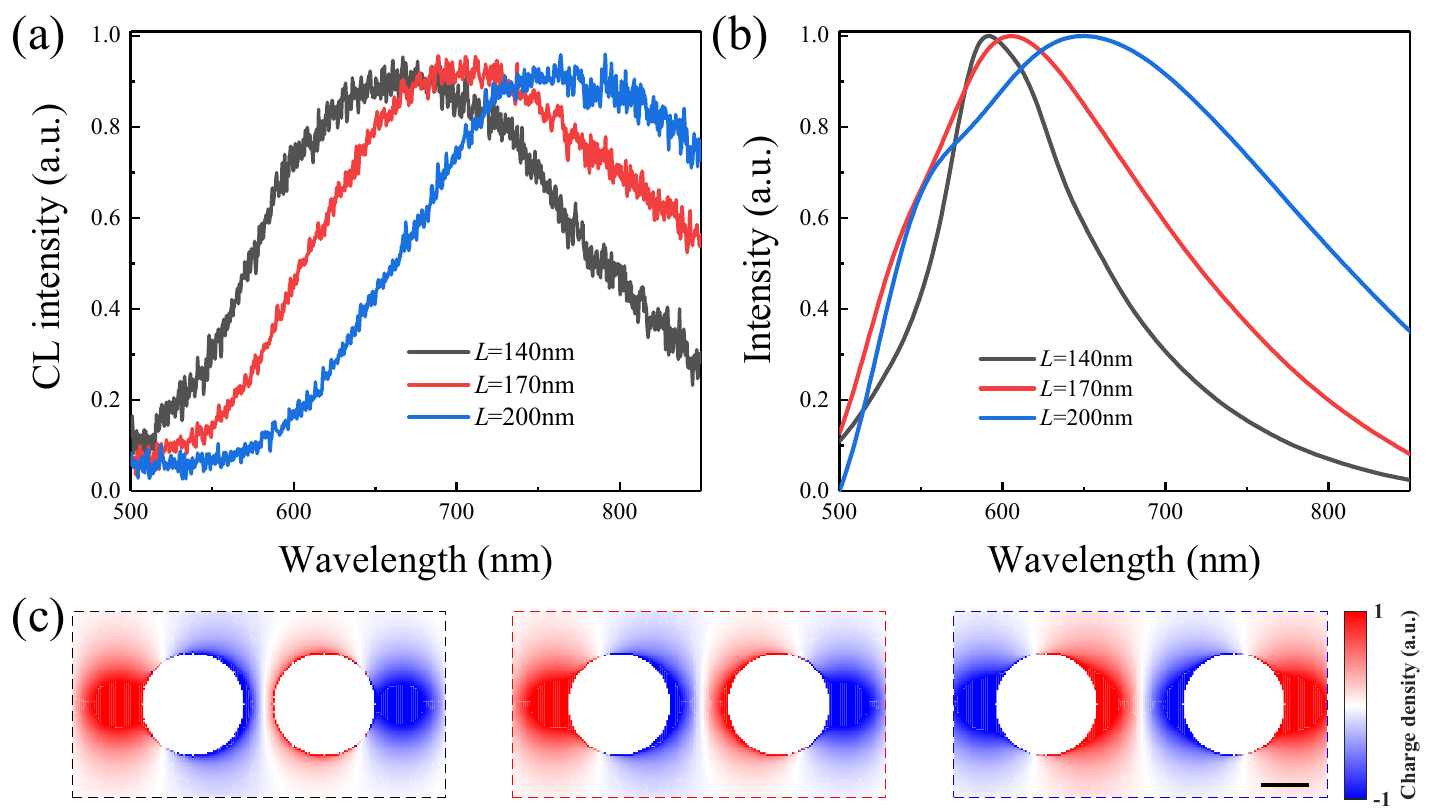}
\caption{Comparison of measured (a) and simulated (b) CL spectra for the nanodimers with separations $L$=140, 170, and 200 nm. (c) The simulated surface charge density distributions of each dimer at the emission peak wavelengths. The scale bar is 50~nm.}\label{fig3}
\end{figure}

On the other hand, for a nanohole dimer with a separation distance well below the resonance wavelength, the near-field coupling between nanoholes becomes significant. In this region, the model of propagated SPP interference breaks down, and the plasmon response of the dimer is determined by its localized plasmon resonance. Figure~\ref{fig3}(a) and (b) indicate the experimental and simulated spectra for the hole dimers with $L$= 140, 170, and 200~nm, a consistent trend between them can be recognized. The resonance wavelength redshifts as the hole separation distance increases, which is attributed to that larger restoring forces are needed for the stronger near-field interaction between holes with small separation distance. Moreover, it has been reported that symmetric and antisymmetric plasmon modes can be supported in a hole dimer \cite{sannomiya2016coupling}. To characterize which resonant mode was excited in our nanohole dimers, the surface charge density on the metal films at peak wavelengths was calculated (Fig.~\ref{fig3}(c)). From the simulations, it is observed that the anti-symmetric mode with opposite charge distribution in the gap is excited by grazing incident electrons.

\begin{figure}[htbp]
\includegraphics[width=120mm,angle=0]{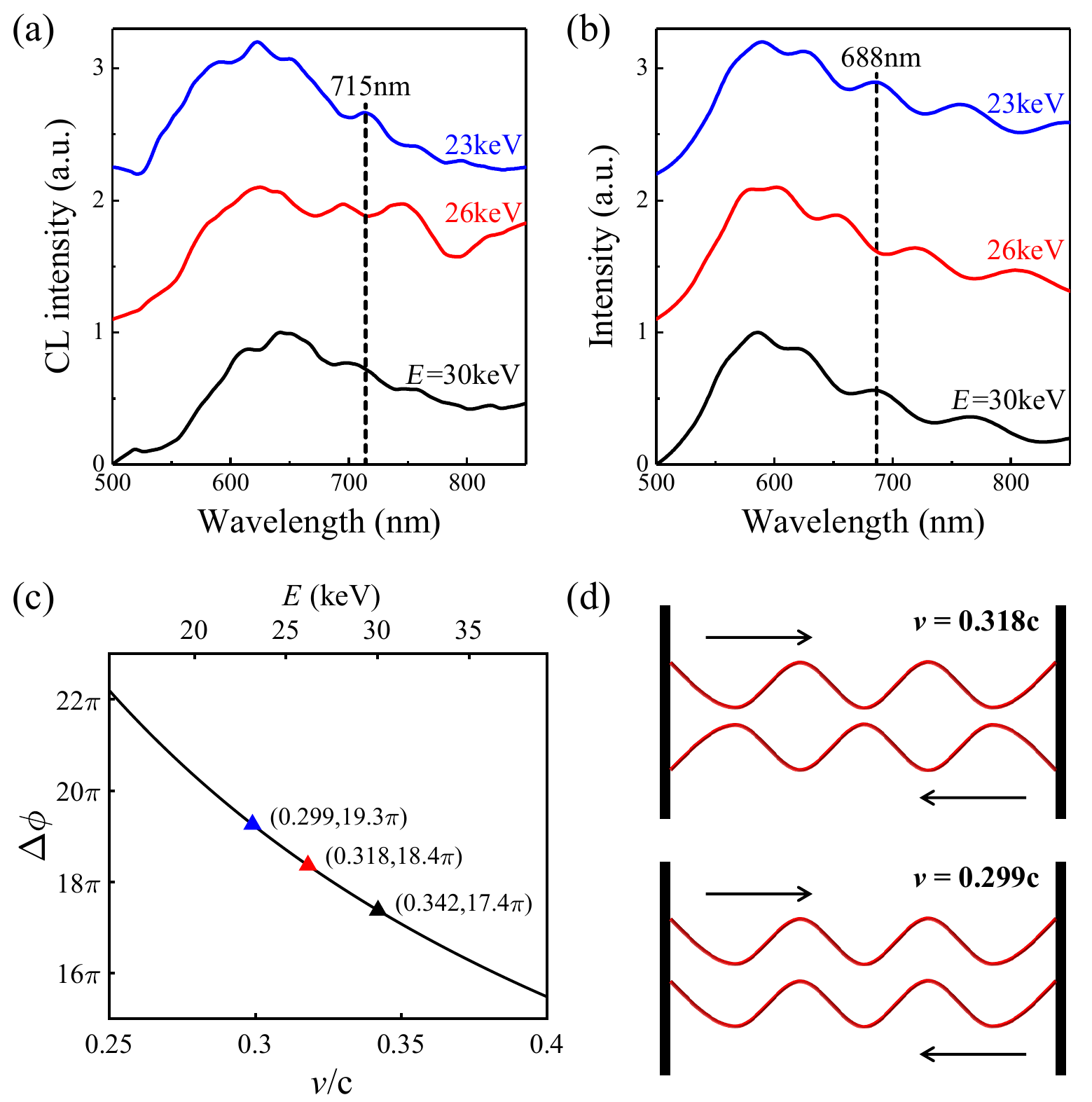}
\caption{(a) Experimental and (b) simulated CL spectra of the nanohole dimer excited by the electron beam with different accelerating voltages 23, 26, and 30 keV with $L$=1600~nm. The dashed lines indicate the peak-valley-peak evolution. (c) The initial phase difference between SPPs launched by the two holes as the variation of the speed of the electron beams. The triangles represent the experimental cases in (a). (d) The schematic of the destructive and constructive interference at 	26 and 23 keV. }\label{fig4}
\end{figure}

Next, compared with previously widely adopted methods for plasmon interference control \cite{morrill2016measuring,dvorak2013control}, another freedom induced by the electron beam can be utilized in our scheme.  Due to the nanometer scale separation distance between holes and the large speed of electrons (for example, 0.34$c$ for 30 keV electron beams), the time delay between SPPs can be controlled within the femtosecond (fs) scale. In addition, for a fixed geometry, the initial phase difference between nanohole launched SPPs can be actively controlled by changing the electron beam energy. So we explored further how the velocity of the electron beam modulates the initial phase difference of SPPs. Figure~\ref{fig4}(a) shows the measured CL spectra for the nanohole dimer with a separation distance of 1600~nm, which was excited by electrons with accelerating voltages of 30 (0.34$c$), 26 (0.32$c$), and 23 kV (0.3$c$), respectively. The peak or valley positions in the spectra change with the increase of accelerating voltage. For example, at the wavelength of 715~nm indicated by dashed lines, a peak-valley-peak evolution exists.  Figure \ref{fig4}(b) shows the corresponding simulations, where a similar evolution happens as well but at 692~nm. The small discrepancy in wavelength can be attributed to the inconsistency of experiments and simulations due to the tapering effects or gallium implantation from FIB milling of the nanoholes. In further, to explain the evolution effect, the initial phase differences of SPPs launched by two holes were calculated through $2\pi c L/v\lambda_0$ ($\lambda_0= 715$~nm) (Fig.~\ref{fig4}(c)). For the experiments, 17.4$\pi$, 18.4$\pi$, and 19.3$\pi$ initial phase differences for the SPPs due to the time delay of electrons with the energy of 30, 26, and 23 keV are obtained, respectively. It is obvious that around a 2$\pi$ phase change was realized during the electron energy control, giving a clear explanation for the peak-valley-peak evolution. In specific, as shown in Fig.~\ref{fig4}(d), the destructive and constructive interference of SPPs are realized at 23 and 26 keV, respectively. Besides, because of the very large speed of electrons, the time duration for the electron passing two holes of the dimer is very small (around 15.7 fs for the 30 kV electron beam). In theory, the time resolution of SPP control ($\Delta t$) with non-relativity electrons can be described as 
\begin{equation}
\Delta t=\sqrt{\frac{m_e}{8}}\frac{L}{E^{3/2}}\Delta E,
\label{eqn2}
\end{equation}
where $m_e$ is the mass of an electron, $E$ is the energy of electrons, and $\Delta E$ is the resolution of electron energy. In our experiments, the estimated time resolution is around 25.9 attosecond (as) at the electron energy of 30 keV with  $\Delta E=0.1$ keV, providing a surprising ultrafast method to control plasmon interference. It is worth mentioning the period of SPPs wave at $\lambda_0$=715~nm is 2.4 fs, to control it with this time is hardly reachable with traditional optical ultrafast laser technique.

\section{CONCLUSION}
In summary, by using electron beams under aloof geometry, the plasmon response of a nanohole dimer was explored systematically. Two different interaction regions were classified depending on the separation distance between holes. In the small separation region, due to the near-field coupling, anti-symmetrical coupled localized plasmon resonance was observed. In contrast, for the large separation distance, the CL emission comes from the interference between hole launched SPPs. Moreover, due to the limited speed of electrons, the initial phase difference between the SPPs generated by holes in the dimer was manipulated, and the CL emission intensity was actively controlled by the electron energy. Meanwhile, the time resolution of this kind of control is shown to be at the attosecond (as) scale, providing an ultrafast tool for plasmon interaction control. Our work not only provides a comprehensive understanding of the plasmon response of a nanohole dimer, but also suggests a novel scheme to control polaritons ultrafastly, which suggests potential applications in ultrafast plasmonics, such as ultrafast optical processing and sensing.

\section{METHODS}
\subsection{Sample preparation.} The magnetron sputtering method was used to deposit a 70 nm thick gold film on a silicon substrate. Subsequently, the circular nanoholes were fabricated in the film using focused ion beam (FIB) milling in an FEI Helios NanoLab 600i system with a Ga$^+$ beam (30 keV, 1.1 pA).

\subsection{CL experiments.} The CL experiments were performed in the scanning electron microscope (SEM) with a parabolic mirror inside. The sample was excited by a grazing incident electron beam with a beam current of 11 nA from the SEM. The impact parameter is about 10~nm. Then the light emitted from the sample was collected by the parabolic mirror and guided to a spectrometer equipped with a liquid-nitrogen-cooled CCD array by an optical fiber. The CL spectra were recorded in the wavelength range of 500-850~nm and corrected by dividing the background signal from the unstructured gold film.

\subsection{Numerical simulation.} Commercial software, Lumerical FDTD Solutions, was used to perform three-dimension numerical simulations. The calculation region is 3~$\mu$m $\times$3~$\mu$m$\times$1.5~$\mu$m, and the size of the gold film is 3000~nm$\times$3000~nm$\times$~70 nm. The diameter of the nanohole dimer is 110~nm. To save the calculation time, the mesh size is set 10 nm$\times$10 nm$\times$5 nm. Perfectly matched layers (PML) boundaries are used to avoid reflection. The electron beam 10 nm above the metal surface is mimicked by a series of $x$-polarized dipoles with phase delay $x/v$, $v$ is the speed of the electron beam. The dielectric function of gold used in the simulations is obtained from Johnson and Christy \cite{johnson1972optical}. The simulated emission spectra were calculated by integrating the $z$ component of the Poynting vector in the upper half-space. The charge density distribution is analyzed by calculating the divergence of the electric displacement vector.

\section*{Acknowledgment}

This work was supported by Guangdong Major Project of  Basic and Applied Basic Research (2020B0301030009), the Program for the National Key R\&D Program of China (2017YFA0305100, 2017YFA0303800), the National Natural Science Foundation of China (91750204, 12074200, 11774185, 12004196, 61775106, 92050114), Changjiang Scholars and Innovative Research Team in University (IRT13\_R29), the 111 Project (B07013), the Tianjin Natural Science Foundation (18JCQNJC02100), and Fundamental Research Funds for the Central Universities.

\providecommand{\newblock}{}

\newpage

\end{document}